\documentclass[12pt,aps,preprint,showpacs]{revtex4}
\usepackage[dvipdfm]{graphicx}
\usepackage{epsfig}
\usepackage{graphicx}
\usepackage{latexsym}
\usepackage{amsmath}
\usepackage{amsfonts}
\usepackage{amssymb}

\begin{document}

\title{Differential Geometry of Polymer Models: Worm-like Chains, Ribbons
and Fourier Knots}
\author{S.M. Rappaport and Y. Rabin }
\affiliation{Department of Physics, Bar--Ilan University, Ramat-Gan
52900, Israel }
\begin{abstract}
We analyze several continuum models of polymers: worm-like chains, ribbons
and Fourier knots. We show that the torsion of worm-like chains diverges and
conclude that such chains can not be described by the Frenet-Serret (FS)
equation of space curves. While the same holds for ribbons as well, their
rate of twist is finite and, therefore, they can be described by the
generalized FS equation of stripes. Finally, Fourier knots have finite
curvature and torsion and, therefore, are sufficiently smooth to be
described by the FS equation of space curves.
\end{abstract}
\maketitle
\section{Introduction}

Recent progress in the ability to manipulate single biomolecules
such as double stranded DNA and protein filaments \cite{Bensimon},
prompted the development of continuum models of complex polymers
capable of describing bending fluctuations and finite extensibility
under tension (extensions of the worm-like chain model \cite{Marko})
as well as models that can describe twist rigidity, spontaneous
twist and chiral response to torque (the ribbon model \cite{rabin}).
These models are defined by their elastic energy functions which are
then used to generate the equilibrium ensemble of polymer
conformations, based on the conventional Gibbs distribution approach
(i.e., weighting the conformations by an appropriate Boltzmann
factor). While this approach proved to be quite successful for open
polymers \cite {mezard,yevgeny}, it could not be applied to generate
the conformations of polymer loops and to study properties of
circular double stranded DNA such as supercoiling and formation of
knots. In order to cope with the latter problem, we developed a
purely mathematical procedure of generating closed curves, based on
the expansion of the components of the polymer conformation vector
$\vec{r}(t)$ ($t$ is some parametrization of the contour of the
loop) in finite Fourier series and taking the corresponding Fourier
coefficients from some random distribution \cite{Shay2}. We found
that this distribution of Fourier knots could be fine tuned to mimic
some of the large scale properties of closed Gaussian loops and
small scale properties of worm-like chain, this could not be
achieved using a single persistence length.

The present deals attempts to establish a common framework for the
discussion of the above physical and mathematical models of
polymers, by classifying them according to their smoothness. A
continuous curve $\vec{r} (s)$ ($s$ is the contour parameter which
measures the distance along the contour) is defined as an $m$-smooth
curve if its $m$th derivative $d^{m} \vec{r}/ds^{m}$ is a continuous
function of $s$. In section \ref{sec:Frenet} we introduce the
fundamental equations of differential geometry of space curves and
of stripes and show that these equations describe space curves that
are at least $2$-smooth and stripes that are at least $1$-smooth.
This is equivalent to saying that while such space curves must have
finite torsion, stripes must have finite rate of twist but their
torsion is free to take any value along their contour. In section
\ref{sec:PhysicalModel} we show that worm-like chains belong to the
class of freely rotating models, with uniformly distributed dihedral
angles, divergent torsion and a normal whose direction jumps
discontinuously as one moves along the contour of the chain and,
therefore, such objects can not be described by the Frenet-Serret
(FS) equation. We also show that because the energy of \ a ribbon
depends on its twist, the typical conformations of \ ribbons have
finite rate of twist but their torsion diverges at many points along
the contour. \ In section \ref{sec:the models} we compare the
ensembles of $1$-smooth (ribbons) and $ \infty $-smooth (Fourier
knots) curves and show that typical realizations of the latter (but
not the former) ensemble, have finite torsion and a smoothly varying
normal and can be described by the FS equation. We also calculate
the distributions of spatial distances between two points on the
contour of the curve in the above ensembles and find that these
distributions differ significantly only for distances of the order
of persistence length. Finally, in section \ref{sec:Discussion} we
discuss our results and conclude that unlike worm-like chains and
ribbons which possess no torsional rigidity, the\ ensemble of curves
generated by the Fourier knot algorithm can be characterized by
finite torsional persistence length.

\section{Differential geometry of curves and stripes}

\label{sec:Frenet} It is often convenient to represent a space curve
$\vec { r }(s)$ defined in a space-fixed coordinate frame by
intrinsic coordinates, as follows. At every point along the curve
($0\leq s\leq L$) one constructs a set of three orthogonal unit
vectors known as the Frenet frame: the tangent $\hat{t}$ defined as
$\hat{t}(s)=d\vec{r}/ds$, the normal $\hat{n} (s) $ which points in
the direction of $d\hat{t}/ds$ and the binormal $\hat{b }(s)=\hat {t
}(s)\times\hat{n}(s)$. The rotation of the Frenet frame as one moves
along the contour of the curve is described by the Frenet-Serret
(FS) equation \cite{willmore}:
\begin{equation}
\frac{d}{ds}\left(
\begin{array}{c}
\hat{t} \\
\hat{n} \\
\hat{b}
\end{array}
\right) =\left(
\begin{array}{ccc}
0 & \kappa & 0 \\
-\kappa & 0 & \tau \\
0 & -\tau & 0
\end{array}
\right) \left(
\begin{array}{c}
\hat{t} \\
\hat{n} \\
\hat{b}
\end{array}
\right)  \label{eq:Frenet equation}
\end{equation}
where $\kappa$ is the curvature and $\tau$ is the torsion (in general, both
are functions of $s$). The condition of validity of the above equation is
that $\kappa ds,\tau ds\rightarrow0$ as $ds\rightarrow0$ everywhere along
the curve. It is straightforward to show that
\begin{equation}
\kappa(s)=\left\vert \frac{d^{2}\vec{r}}{ds^{2}}\right\vert ,\qquad \tau(s)=
\frac{d\vec{r}}{ds}\cdot\left( \frac{d^{2}\vec{r}}{ds^{2}}\times \frac{d^{3}
\vec{r}}{ds^{3}}\right) \left\vert \frac{d^{2}\vec{r}}{ds^{2} }\right\vert
^{-2}  \label{curvtor}
\end{equation}
and we conclude that since the condition of validity of the FS equation is
that the torsion is finite everywhere along the curve, the curve should be
at least $2$-smooth.

Given the curvature and the torsion at each point along the curve,
one can solve Eq. (\ref{eq:Frenet equation}), calculate the tangent
$\hat{t}(s)$ and integrate it to construct the parametric
representation of the space curve, $ \vec{r}(s)$. The above
construction is unique in the sense that any pair of functions
$\kappa (s)$ and $\tau (s)$ defined on the interval $0<s<L$, can be
uniquely mapped to a space curve $\vec{r}(s)$ of length $L$. The
simplest examples are (a) $\kappa =const.,\tau =0$ which yields a
planar circle of radius $\kappa ^{-1}$ and (b) $\kappa =const.,\tau
=const.$ which corresponds to a helix.

We now turn to consider stripes of length $L$, width $W$ (such that
$L\gg W$ ) and thickness $D\rightarrow 0$. Unlike a space curve
which is uniquely defined by the tangent vector $\hat{t}(s)$ (the
normal and the binormal are auxiliary constructs, needed only to
calculate the tangent, given the curvature and the torsion), a
stripe is a slice of a plane with which one can associate two
orthogonal unit vectors $\hat{t}_{1}$ (in plane) and $\hat{ t}_{2}$
(normal to the plane). The spatial configuration of the stripe is
thus defined by the local orientation (at each point $s$ on the
centerline) of the orthogonal triad known as the Darboux frame which
specifies the directions of the two axes $\hat{t}_{1}$ and
$\hat{t}_{2}\,$ and that of the tangent to the centerline that runs
along the long axis of the stripe, $\hat{ t}\equiv \hat{t}_{3}$.
While a space curve is completely defined by the two functions
$\kappa (s)$ and $\tau (s)$, a stripe is represented by three
generalized curvatures $\omega _{k}(s)$ ($k=1,2,3$) that determine
the unit vectors $\left\{ \hat{t}_{i}\right\} $ via the generalized
Frenet-Serret equation \cite{SolidShape},
\begin{equation}
\frac{d}{ds}\left(
\begin{array}{c}
\hat{t}_{3} \\
\hat{t}_{1} \\
\hat{t}_{2}
\end{array}
\right) =\left(
\begin{array}{ccc}
0 & \omega _{2} & -\omega _{1} \\
-\omega _{2} & 0 & \omega _{3} \\
\omega _{1} & -\omega _{3} & 0
\end{array}
\right) \left(
\begin{array}{c}
\hat{t}_{3} \\
\hat{t}_{1} \\
\hat{t}_{2}
\end{array}
\right) .  \label{eq:Ribbon equation}
\end{equation}
This can be written compactly as
$\dot{\hat{t}}_{i}=A_{ik}\hat{t}_{k}$ where $A_{ik}=-\epsilon
_{ijk}\omega _{j}$ ($\epsilon _{ijk}$ is the Levi-Civita tensor).
Inspection of Eq. (\ref{eq:Ribbon equation}) shows that $\omega
_{i}(s)ds$ is the infinitesimal angle of rotation about the
direction $\hat{t }_{i}(s)$ and the condition of validity of the
generalized FS equation is that this angle vanishes in the limit
$ds\rightarrow 0$. However, unlike the torsion $\tau \,$\ which is
completely determined by the space curve $\vec{r} (s)$ and can be
expressed in terms of its first three derivatives, the rate of twist
$\omega _{3}$ is the local rate (per unit length) of rotation about
the tangent to this curve and, as such, it depends only on the
orientation of $\hat{t_{1}}$ and $\hat{t_{2}},$ and can not be
expressed in terms of the centerline $\vec{r}(s)$ and its
derivatives! We conclude that in order for Eq. (\ref{eq:Ribbon
equation}) to hold, the centerline of the stripe should be
represented by a\thinspace $1$-smooth curve.

Since both couples of unit vectors $\left\{ \hat{n},\hat{b}\right\}
$ and $ \left\{ \hat{t}_{1},\hat{t}_{2}\right\} $ lie in the plane
perpendicular to the local tangent, the triads $\left\{
\hat{t},\hat{n},\hat{b}\right\} $ and $\left\{
\hat{t_{3}},\hat{t}_{1},\hat{t}_{2}\right\} $ are connected by
\begin{equation}
\left(
\begin{array}{c}
\hat{t}_{3} \\
\hat{t}_{1} \\
\hat{t}_{2}
\end{array}
\right) =\overleftrightarrow{S}\left(
\begin{array}{c}
\hat{t} \\
\hat{n} \\
\hat{b}
\end{array}
\right)   \label{eq:FrRib}
\end{equation}
where the matrix
\begin{equation}
\overleftrightarrow{S}=\left(
\begin{array}{ccc}
1 & 0 & 0 \\
0 & \cos \alpha  & \sin \alpha  \\
0 & -\sin \alpha  & \cos \alpha
\end{array}
\right)   \label{eq:S}
\end{equation}
generates a rotation by an angle $\alpha $ about the
$\hat{t}=\hat{t}_{3}$ axis. Assuming that the centerline of the
stripe can be represented by the FS equation (i.e., that it is an
$2$-smooth curve), one can express the generalized FS equation in
terms of the curvature, torsion and the angle $ \alpha $ between the
directions of the binormal $\hat{b}$ and the $\hat{t_{1} }$\ axis:
\begin{equation}
\frac{d}{ds}\left(
\begin{array}{c}
\hat{t}_{3} \\
\hat{t}_{1} \\
\hat{t}_{2}
\end{array}
\right) =\left(
\begin{array}{ccc}
0 & \kappa \cos \alpha  & -\kappa \sin \alpha  \\
-\kappa \cos \alpha  & 0 & \dfrac{d\alpha }{ds}+\tau  \\
\kappa \sin \alpha  & -\dfrac{d\alpha }{ds}-\tau  & 0
\end{array}
\right) \left(
\begin{array}{c}
\hat{t}_{3} \\
\hat{t}_{1} \\
\hat{t}_{2}
\end{array}
\right) .  \label{eq:RibbEqu}
\end{equation}
Comparing (\ref{eq:RibbEqu}) with (\ref{eq:Ribbon equation}) yields:
\begin{equation}
\begin{array}{c}
\omega _{1}=\kappa \sin (\alpha ) \\
\omega _{2}=\kappa \cos (\alpha ) \\
\omega _{3}=d\alpha /ds+\tau
\end{array}
\label{eq:omega}
\end{equation}
Conversely,
\begin{equation}
\kappa =\sqrt{\omega _{1}^{2}+\omega _{2}^{2}}  \label{eq:Kap}
\end{equation}
\begin{equation}
d\alpha /ds=\frac{\omega _{2}\left( d\omega _{1}/ds\right) -\omega
_{1}\left( d\omega _{2}/ds\right) }{\omega _{1}^{2}+\omega _{2}^{2}}
\label{eq:Alph}
\end{equation}
\begin{equation}
\tau =\omega _{3}-d\alpha /ds  \label{eq:Tau}
\end{equation}
However, since the centerline of the stripe is required to be only $1$
-smooth, these relations do not hold in general!

\section{Polymer models: Worm-like Chains and Ribbons}

\label{sec:PhysicalModel}Physical models of polymers are based on
the choice of a geometrical model and an energy functional. The
simplest and the most prevalent model is that of a continuous
Gaussian random walk, with which one can associate a free energy
that describes the entropic cost of stretching the polymer chain,
$E_{GRW}=\frac{1}{2}ak_{B}T\int\nolimits_{0}^{L}{dt} \left(
{d}\vec{{r}}{/dt}\right) ^{2}.$ Here $a$ is the \textquotedblleft
monomer\textquotedblright\ (cutoff) length, $k_{B}$ is the Boltzmann
constant and $T$ is the temperature \cite{colby}. Notice that this
free energy is expressed only in terms of the first derivative of
the trajectory, ${d\vec{{r}}/dt}$ and, therefore, space curves that
describe polymer conformations in the continuous Gaussian random
walk model have to be only $0 $-smooth (only the curve itself and
not its derivatives, has to be continuous everywhere). \newline The
worm-like chain model of polymers combines bending elasticity and
inextensibility (the latter condition can be expressed as
$\left\vert {d\vec{ {r}}/ds}\right\vert =1$) and, assuming that the
stress-free state corresponds to a straight line, the energy can be
written as \cite{Kamien}
\begin{equation}
E_{WLC}=\frac{1}{2}b\int\nolimits_{0}^{L}{ds(\kappa
(s))^{2}=}\frac{1}{2} b\int\nolimits_{0}^{L}{ds(d\hat{t}/ds)^{2}}.
\label{Ewlc}
\end{equation}
Since the energy depends only on the curvature ${\kappa }$ and does
not depend on the torsion $\tau ,$ the corresponding space curve has
only $1$ -smooth. Note that curves described by the FS equations
have to be at least $ 2$-smooth and, therefore, worm-like chains are
not sufficiently smooth to be described by the fundamental equations
of differential geometry of space curves! In order to get physical
intuition about the origin of the problem, lets consider a
discretized model of a continuous curve in which the polymer is made
up of connected straight segments of length $\Delta s$ each, such
that the direction of the segment at point $s$ is given by the
tangent to the original chain at this point, $\hat{t}(s)$ (the
continuum limit is recovered as $\Delta s\rightarrow 0$). The angle
between neighboring segments is denoted as $\Delta \theta (s)$ and,
in order to describe the non-planar character of a general space
curve, one has to introduce the dihedral angle $\Delta \varphi (s)$
between the two successive planes $ \left\{
\hat{t}(s),\hat{t}(s+\Delta s)\right\} $ and $\left\{ \hat{t}
(s+\Delta s),\hat{t}(s+2\Delta s)\right\} $ determined by three
successive segments at points $s,$ $s+\Delta s$ and $s+2\Delta s$.
Since $\Delta \varphi (s)$ is also the angle between neighboring
binormals $\hat{b}(s)$ and $\hat{b}(s+\Delta s)$, the Frenet frames
at points $s$ and $s+\Delta s$ are related by a simple rotation
\begin{equation}
\left(
\begin{array}{c}
\hat{t}(s+\Delta s) \\
\hat{n}(s+\Delta s) \\
\hat{b}(s+\Delta s)
\end{array}
\right) =\overleftrightarrow{B}(s)\left(
\begin{array}{c}
\hat{t}(s) \\
\hat{n}(s) \\
\hat{b}(s)
\end{array}
\right)   \label{eq:discreet equation}
\end{equation}
where the rotation matrix $\overleftrightarrow{B}(s)$ is given by
\begin{equation}
\overleftrightarrow{B}(s)=\left(
\begin{array}{ccc}
\cos \Delta \theta  & \sin \Delta \theta  & 0 \\
-\cos \Delta \varphi \sin \Delta \theta  & \cos \Delta \varphi \cos \Delta
\theta  & \sin \Delta \varphi  \\
\sin \Delta \varphi \sin \Delta \theta  & -\sin \Delta \varphi \cos \Delta
\theta  & \cos \Delta \varphi
\end{array}
\right)   \label{eq:B}
\end{equation}
Notice that no assumption is made so far about the magnitude of the
angles $ \Delta \theta $ and $\Delta \varphi $. Defining $\Delta
\hat{t}(s)=\hat{t} (s+\Delta s)-\hat{t}(s),$ $\Delta $
$\hat{n}(s)=\hat{n}(s+\Delta s)-\hat{n} (s)$ and $\Delta
\hat{b}(s)=\hat{b}(s+\Delta s)-\hat{b}(s)$ and subtracting the
vector $(\hat{t},\hat{n},\hat{b})$ from both sides of Eq.
(\ref{eq:B}), \ yields
\begin{equation}
\left(
\begin{array}{c}
\Delta \hat{t} \\
\Delta \hat{n} \\
\Delta \hat{b}
\end{array}
\right) =\left( \overleftrightarrow{B}(s)-\overleftrightarrow{I}\right)
\left(
\begin{array}{c}
\hat{t}(s) \\
\hat{n}(s) \\
\hat{b}(s)
\end{array}
\right)   \label{eq:discrete frenet equation}
\end{equation}
with $\overleftrightarrow{I}$ the unit matrix. Notice that unlike
the FS equation which is valid only for infinitesimal rotations of
the Frenet frame, Eq. (\ref{eq:discrete frenet equation}) describes
finite rotations; the FS equation can be derived from it by dividing
both sides of the equation by $\Delta s$ and taking the limit
$\Delta s\rightarrow 0$ ($ \underset{\Delta s\rightarrow 0}{\lim
}\Delta \hat{t}/\Delta s=d\hat{t}/ds$, etc.). In order for the right
hand side of Eq. (\ref{eq:discrete frenet equation}) to remain
finite in this limit, all the elements of $
\overleftrightarrow{B}(s)-\overleftrightarrow{I}$ have to vanish
$.$Since this is equivalent to the condition $\Delta \theta ,$
$\Delta \varphi \ll 1$ , one can expand the cosine and sine
functions in Eq. ( \ref{eq:B}) and, upon substituting the result
into Eq. (\ref{eq:discrete frenet equation}), one recovers the FS
equation with $\kappa =\lim\limits_{\Delta s\rightarrow 0}\Delta
\theta /\Delta s$ and $\tau =\lim\limits_{\Delta s\rightarrow
0}\Delta \varphi /\Delta s.$

\begin{figure}[ptb]
\centerline{(a)\epsfig{file=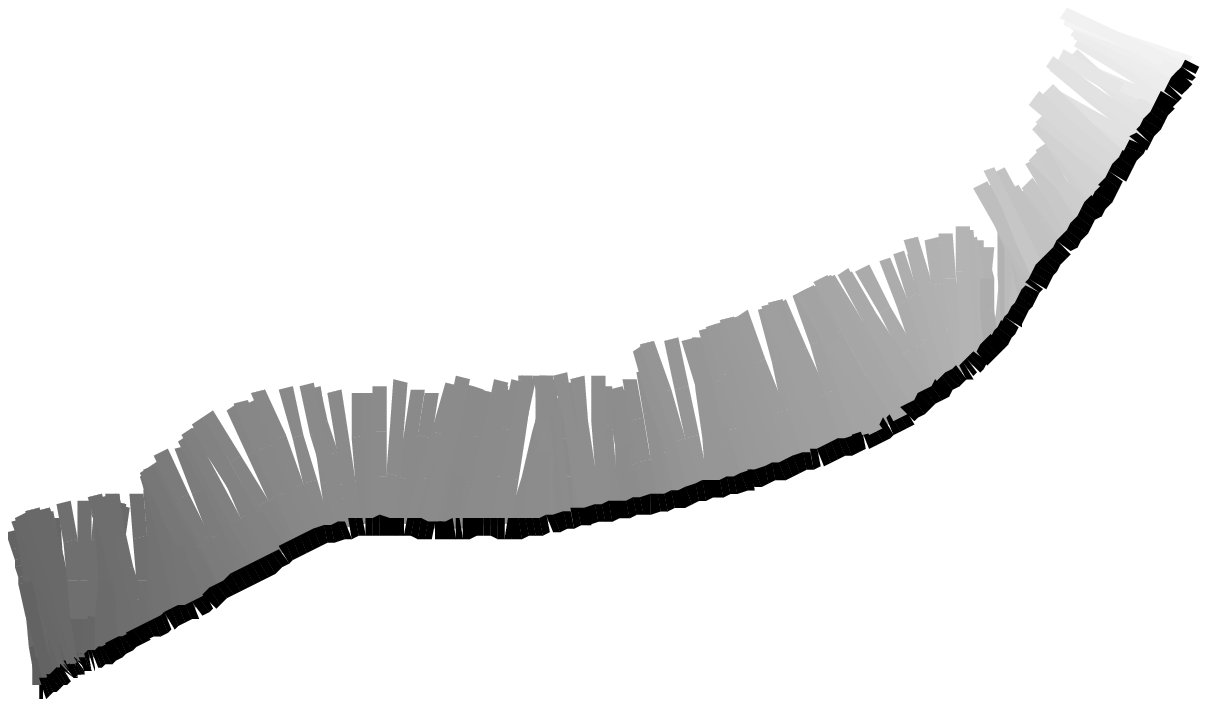,width=0.35\linewidth,angle=-70}(b)
\epsfig{file=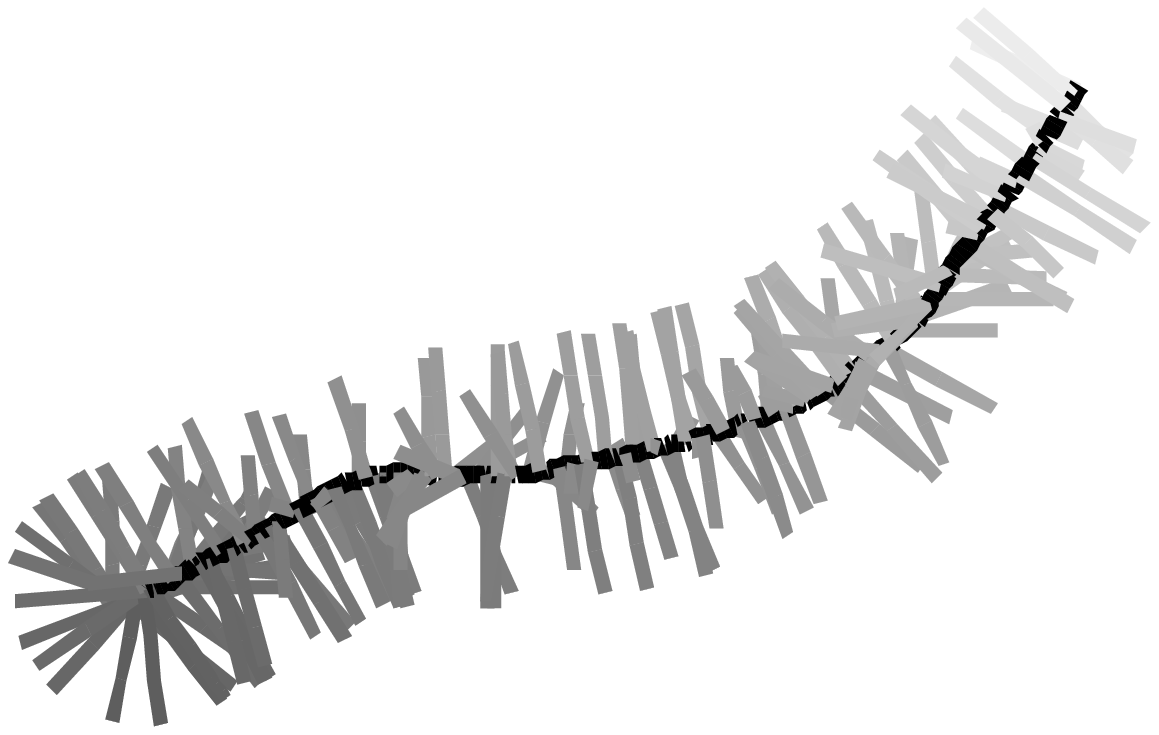,width=0.35\linewidth,angle=-70}}
\caption{A typical conformation of a ribbon: the centerline is shown by the
dark solid curve and the gray lines orthogonal to it show the direction of
(a) one of the symmetry axes of the cross section $\hat{t}_{1}$ and (b) the
normal $\hat{n}$ }
\label{fig:fig1}
\end{figure}

Returning to the worm-like chain model, we notice that while the
bending energy ensures that the curvature is finite and the angle
$\Delta \theta $ is always small, there is no corresponding physical
restriction on the magnitude of $\Delta \varphi $ and we conclude
that the worm-like chain corresponds to the class of freely rotating
chain models in which the angle $ \Delta \varphi $ can attain any
value in the interval $\left[ -\pi ,\pi \right] $ . For such models,
the expansion of $\overleftrightarrow{B}(s)- \overleftrightarrow{I}$
in terms of $\Delta \varphi $ breaks down and the corresponding
($1$-smooth) curves can not be described by the FS equation. Notice
that if one keeps the definition $\tau =\lim\limits_{\Delta
s\rightarrow 0}\Delta \varphi /\Delta s,$ the torsion can diverge at
any point along the curve, generating abrupt jumps of the normal
(see Fig. \ref {fig:fig1} b). Nevertheless, since the shape of a
curve is completely characterized by (and only by) the tangent to it
and since the latter changes continuously even for $1$-smooth
curves, such a curve appears (to the eye) to be just as smooth  as
an $\infty $-smooth one.

Let us now consider the ribbon model of polymers which was designed
to take into account rigidity with respect to twist and spontaneous
twist of complex polymers such as double stranded DNA. In general,
the ribbon has an asymmetric cross section with symmetry axes
$\hat{t_{1}}$ and $\hat{t_{2}}$ and a centerline described by the
tangent, $\hat{t_{3}}$, such that the triad of unit vectors $\left\{
\hat{t_{i}}(s)\right\} \ $can be associated with the Darboux frame
familiar from the differential geometry of stripes. In the framework
of the linear theory of elasticity of slender rods \cite {Love}),
the energy of a particular configuration of a ribbon is a quadratic
functional of the deviations of its three curvatures $\left\{ \omega
_{i}(s)\right\} $ from their equilibrium values in the stress-free
state, $ \left\{ \omega _{0i}(s)\right\} $ \cite{rabin}:
\begin{equation}
\begin{array}{c}
E_{R}=\frac{1}{2}\int\nolimits_{0}^{L}ds\left[ b_{1}\left( \omega
_{1}-\omega _{01}\right) ^{2}+b_{2}\left( \omega _{2}-\omega _{02}\right)
^{2}\right.  \\
+\left. b_{3}\left( \omega _{3}-\omega _{03}\right) ^{2}\right]
\end{array}
\end{equation}
Here $b_{1}$ and $b_{2}$ are the bending rigidities associated with the two
principal symmetry axes of the cross section and $b_{3}$ is the twist
rigidity (the persistence lengths $\left\{ a_{i}\right\} $ are obtained by
dividing the corresponding rigidities by $k_{B}T$). For ribbons with a
symmetric cross-section ($b_{1}=b_{2}$) and without spontaneous curvature
and twist ($\omega _{0i}=0$), the above expression can be simplified
\begin{equation}
E_{R}=\frac{1}{2}\int\nolimits_{0}^{L}ds\left( b_{1}\kappa ^{2}+b_{3}\omega
_{3}^{2}\right)   \label{Eribbon}
\end{equation}
where $\kappa ^{2}=$ $\omega _{1}^{2}+\omega _{2}^{2}$. Since the
above energy functional depends only on the second derivatives of
the curve (curvature) and on the twist, and does not explicitly
depend on the third derivatives (torsion), the torsion of the curve
is not controlled by the elastic energy and, therefore, the
centerline of the ribbon can not be described by the FS equation. To
demonstrate it, we rewrite Eq.(\ref {eq:Ribbon equation}) in a
discretized form, using (\ref{eq:B}) and the relation (\ref{eq:S}):
\begin{equation}
\left(
\begin{array}{c}
\Delta \hat{t_{3}} \\
\Delta \hat{t_{1}} \\
\Delta \hat{t_{2}}
\end{array}
\right) =\left( \overleftrightarrow{S}(s+\Delta
s)\overleftrightarrow{B}(s)
\overleftrightarrow{S}^{-1}(s)-\overleftrightarrow{I}\right) \left(
\begin{array}{c}
\hat{t_{3}} \\
\hat{t_{1}} \\
\hat{t_{2}}
\end{array}
\right) .  \label{eq:discrete RibEq}
\end{equation}
Comparing (\ref{eq:discrete RibEq}) to (\ref{eq:Ribbon equation}) gives, in
the limit $\Delta s\rightarrow 0$:
\begin{equation}
\begin{array}{c}
\omega _{3}=\Delta \alpha /\Delta s+\Delta \varphi /\Delta s\rightarrow
d\alpha /ds+\tau  \\
\omega _{1}=\left( \Delta \theta /\Delta s\right) \sin (\alpha )\rightarrow
\kappa \sin \alpha  \\
\omega _{2}=\left( \Delta \theta /\Delta s\right) \cos (\alpha )\rightarrow
\kappa \cos \alpha
\end{array}
\label{eq:discrete omega}
\end{equation}
The fact that the elastic energy of a ribbon introduces a penalty
for bending and twist deformations, ensures that only configurations
with $ \omega _{i}\Delta s\rightarrow 0$ ($i=1,2,3$)\ contribute in
the continuum limit $\Delta s\rightarrow 0$ and, therefore, the
conformations of a ribbon can be described by the generalized FS
equation ( \ref{eq:Ribbon equation}). Inspection of Eq.
(\ref{eq:discrete omega}) shows that the condition that the twist
accumulated over a contour distance $\Delta s,$ is small,  can be
expressed as $\Delta \alpha +\Delta \varphi <<1$. Since there is
nothing that restricts the magnitudes of $\Delta \alpha $ and of
$\Delta \varphi $ separately, they can be arbitrarily large provided
that the condition $ \Delta \alpha \simeq -\Delta \varphi $ is
satisfied and, therefore, the torsion associated with the centerline
of the ribbon can be arbitrarily large (recall that the torsion is
defined as the limit of $\Delta \varphi /\Delta s$ as $\Delta
s\rightarrow 0$). Indeed, inspection of a typical conformation of a
ribbon (taken from the ensemble of conformations generated using the
algorithm described in ref. \cite{yevgeny}) shows that even though
the rate of rotation of the physical axis $\hat{t_{1}}$ is
everywhere finite (Fig. \ref{fig:fig1}a), the rate of rotation of
the normal $\hat{n}$ is not (Fig. \ref{fig:fig1}b)!\\

\section{Ensembles of $1$-smooth and $\infty$-smooth Curves}

\label{sec:the models} 
\begin{figure}[ptb]
\centerline{\scalebox{0.9}{\includegraphics{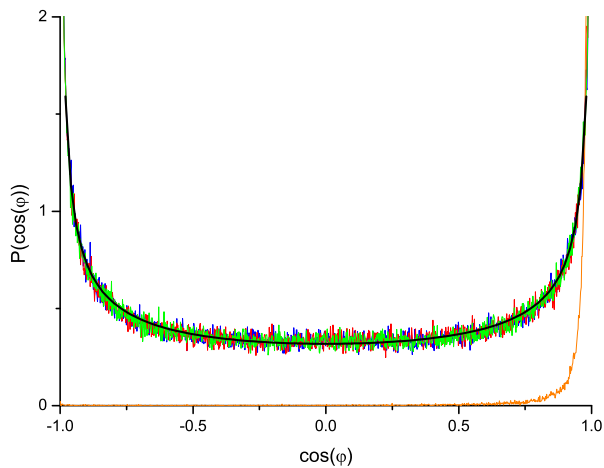}}}
\caption{(color online) The distribution function of
$\cos(\protect\varphi) $ of a worm-like chain (black solid line) and
of a symmetric ribbon with $ a_{1} =a_{2}=0.1$, $a_{3}=0.01$ (red);
$a_{1}=a_{2}=0.1$, $a_{3}=1000$ (green); $a_{1}=a_{2}=0.01$,
$a_{3}=1000$ (blue). The distribution of a Fourier knot with
$n_{0}=150$ is shown by the orange line.} \label{fig:cosphi}
\end{figure}

\begin{figure}[ptb]
\centerline{\scalebox{0.9}{\includegraphics{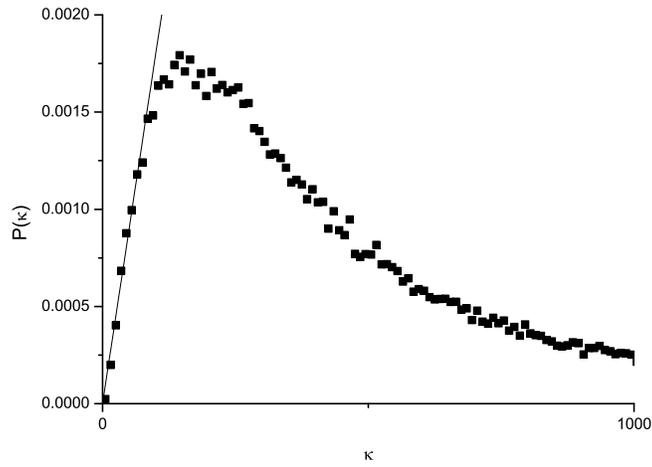}}}
\caption{The distribution of curvature of Fourier knots. The solid line is a
linear fit to the distribution at $\protect\kappa\rightarrow 0$.}
\label{fig:kappa distribution}
\end{figure}

Increasing the degree of smoothness from $m$ to $m+1$
acts as a constraint that prohibits certain configurations of a curve, and
it is interesting to compare the properties of curves with different degrees
of smoothness. Such a comparison is meaningful only in a statistical sense
and in the following we will consider some physically relevant statistical
properties of worm-like chains for which analytical results are available,
with those of computer generated ensembles of ribbons ($1$-smooth)\ and
Fourier knots ($\infty $-smooth). In order to generate the ensemble of
centerlines of ribbons ($1$ -smooth curves), we use the so called
\textquotedblleft Frenet algorithm\textquotedblright\ described in detail in
ref. \cite{yevgeny}. In view of the discussion in the preceding section,
such curves can not be described by the FS equations and, in order to avoid
possible misinterpretation, we will refer to it as the ribbon algorithm.
\newline
Let us compare the distribution of the angle $\Delta \varphi $
between neighboring binormals (or rather of $\cos \Delta \varphi
=\hat{b}(s)\cdot \hat{b}(s+\Delta s)$) in the discretized version of
the worm-like chain model, with that obtained by generating the
ensemble of ribbon conformations, computing the centerline of each
conformation and extracting the distribution of $\cos (\Delta
\varphi )$. In the worm-like chain model $ \Delta \varphi $ is
distributed uniformly in the interval $[-\pi ,\pi ]$ and, therefore,
the probability distribution of $\cos (\Delta \varphi )$ is given by
\begin{equation}
P_{WLC}\left( \cos \Delta \varphi \right) =\frac{1}{\pi \sqrt{1-\cos
^{2}\Delta \varphi }}d\left( \cos \Delta \varphi \right)
\label{eq:cosdistribution}
\end{equation}
Using the ribbon algorithm to obtain the ensemble of conformations
of a ribbon with a symmetric cross section and without spontaneous
curvature, we generate the corresponding distribution $P_{R}$
$\left[ \cos (\Delta \varphi )\right] $. Up to numerical accuracy we
find that the above distribution coincides with the simple worm-like
chain expression, Eq. \ref {eq:cosdistribution} and does not depend
on the bending or twist rigidity (see Fig. \ref{fig:cosphi}). This
concurs with our expectation that, just like worm-like chains,
centerlines of ribbons can be described by freely rotating type
models.\newline In order to generate $\infty $-smooth curves we use
the Fourier knot algorithm which was originally developed with the
goal of investigating knots (i.e., closed curves) \cite{Shay2}.
Unlike methods based on modeling the knot as a $0$-smooth curve made
of discrete, freely jointed segments \cite{Koniaris,Deguchi,natan}),
this algorithm generates infinitely smooth knots, such that the
derivatives $\left\vert d^{m}\vec{r}/dt^{m}\right\vert $ are finite
for all $m$. The Fourier knot algorithm is based on the fact that
for any closed curve parameterized by some arbitrary parameter $t$,
the projections of the position vector $\vec{r}(t)$ on the Cartesian
coordinate axes are periodic functions $r_{i}(t)=r_{i}(t+T)$ with
period $T$ and can be expressed as finite Fourier sums:
\begin{equation}
r_{i}(t)=\sum_{n=1}^{n_{\max }}\left[ A_{n}^{i}\cos (\frac{2\pi
nt}{T} )+B_{n}^{i}\sin (\frac{2\pi nt}{T})\right]
\label{eq:Fourier knot}
\end{equation}
Different realizations of closed curves can be generated by choosing
coefficients $A_{n}^{i},B_{n}^{i}$ from some statistical
distribution. When the coefficients are given by $\lambda n^{-1}\exp
{-n/n_{0}}$, where $ \lambda $ are random numbers in the interval
$\left[ -1,1\right] $ and $n_{0} $ is an effective cutoff ($n_{0}\ll
n_{\max }$), the long wavelength properties of the ensemble
generated by the Fourier knot algorithm, are in good agreement with
those obtained from the worm-like chain model. These properties
include second moments such as the mean square distance between two
points on the contour $\left\langle \left[
\vec{r}(s_{1})-\vec{r}(s_{2}) \right] ^{2}\right\rangle $ \ and the
tangent auto-correlation function $ \left\langle \left[
\vec{t}(s_{1})\cdot \vec{t}(s_{2})\right] \right\rangle $ where
$\left\vert s_{1}-s_{2}\right\vert >>$ $l,$ with the persistence
length $l$ determined by the cutoff as $l=0.58L/n_{0}$. However,
even though the tangent auto-correlation function decays
exponentially with $\left\vert s_{1}-s_{2}\right\vert $ on length
scales comparable to $l$ (just like in the worm-like chain model),
the corresponding decay length $ l_{d}=0.435L/n_{0}\ $is smaller
than the persistence length obtained from the long-wavelength
properties of Fourier knots. In ref. \cite{Shay2} we suggested that
the ensemble of configurations generated by the Fourier knot
algorithm is equivalent to a physical ensemble of polymers which
possess both bending and twist rigidity and, while the short range
properties of the tangent-tangent correlation function are
determined by the bending persistence length only, both bending and
twist persistence length control its long distance behavior. In any
case, the fact that the ensemble of Fourier knots can not be
characterized by a single persistence length suggests that the
statistical properties of this ensemble differ from those of
worm-like chains and that it is important to investigate not only
the second moments but the entire distributions. \newline The first
property we examine is the probability distribution $P_{FK}\left(
\cos \Delta \varphi \right) $. As can be seen in Fig.
\ref{fig:cosphi}, the distribution has a peak at $\cos \Delta
\varphi =1,$ i.e., at $\Delta \varphi =0$. Since $\Delta \varphi
=\tau \Delta s$, \ we conclude that the ensemble of curves generated
by the Fourier knot algorithm is characterized by finite torsion and
a normal whose direction varies smoothly along the contour of each
curve and, therefore, such curves can be described by the FS
equation. We would like to stress that even though the torsion is
described by the first three derivatives of $\vec{r}$ all of which
are finite for Fourier knots, the observation that the ensemble of
Fourier knots is dominated by curves with finite torsion is
non-trivial since the expression for the torsion diverges at points
along the contour where the curvature vanishes (see Eq.
\ref{curvtor}). Notice that for ribbons with no spontaneous
curvature and twist, the partition function can be written as the
product of bending and twist parts $Z=Z_{bend}Z_{twist},$ with
$Z_{bend}$ is given by the functional integral (assuming a symmetric
ribbon of bending persistence length $l$) \cite{rabin}
\begin{align}
Z_{bend}& =\int D\left\{ \omega _{1}\right\} \int D\left\{ \omega
_{2}\right\} e^{-l/2\int ds\left( \omega _{1}^{2}+\omega _{2}^{2}\right) }
\label{Z} \\
& =\int D\left\{ \kappa \right\} \kappa e^{-l/2\int ds\kappa ^{2}}.  \notag
\end{align}
Since $\omega _{1}^{2}+\omega _{2}^{2}=\kappa ^{2}$, the measure
$D\left\{ \omega _{1}\right\} D\left\{ \omega _{2}\right\} $ can be
written as the product of a \textquotedblleft
radial\textquotedblright\ contribution $ D\left\{ \kappa \right\}
\kappa \,$and an angular one. We therefore conclude that the
probability of points with $\kappa \rightarrow 0$ vanishes linearly
with $\kappa $ and since $\tau \propto 1/\kappa ^{2}$, the torsion
should be finite everywhere, as observed. Strictly speaking the
above argument was derived for open ribbons and not to closed
curves, but since it involves only the measure and not the form of
the energy function, it applies to Fourier knots as well  (see Fig.
\ref{fig:kappa distribution} where the measured distribution is
plotted for Fourier knots with $n_{0}=150$).

\begin{figure}[ptb]
\centerline{\scalebox{0.9}{\includegraphics{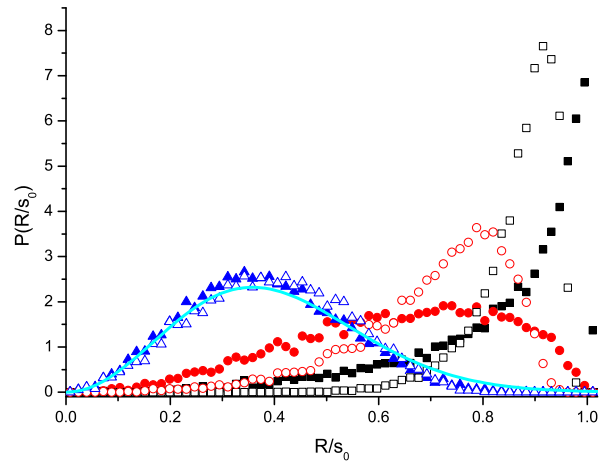}}}.
\caption{(color online) $P(R|s_{0})$ vs $R/s_{0}$ is plotted for
Fourier knots (full symbols) and symmetric ribbons (empty symbols),
of total length $ L=4\protect\pi$ and contour distance
$s_{0}=L/100$. Taking the persistence length of the ribbon as
$l=0.58\ast L/n_{0}$ we plot the distribution for: $ n_{0}=50$,
$l/s_{0}=1.16$ (black squares); $n_{0}=150$, $l/s_{0}=0.386$ (red
circles); $n_{0}=600$, $l/s_{0}=0.096$ (blue triangles). The
Gaussian distribution with $l/s_{0}=0.096$ is shown by the solid
cyan line.} \label{fig:RMS}
\end{figure}

We now turn to compare the statistical properties of $\infty$-smooth
curves generated by the Fourier knot algorithm and the $1$-smooth
centerlines of ribbons generated by the ribbon algorithm. Consider
the probability distribution $P(R|s_{0})$ of the distance
$R(s_{0})=\left\vert \vec{r} (s_{1})-\vec{r}(s_{2})\right\vert $
between points $s_{1}$ and $s_{2}$ ($ s_{0}\equiv\left\vert
s_{1}-s_{2}\right\vert $) along the contour (the second moment of
this distribution for Fourier knots was calculated in ref.
\cite{Shay2}). The difficulty in comparing the two ensembles is that
while the ribbon algorithm generates open curves, the Fourier knot
algorithm yields closed loops. In order to compare the latter with
the former, we make use of the fact that, as long as we consider
contour distances ($s_{0})$ much shorter than the total length of
the loop ($L$), $P_{FK}(R|s_{0})$ approaches the probability
distribution for an open, infinitely smooth curve. In Fig.
\ref{fig:RMS} we plot $P_{R}(R|s_{0})$ (ribbon) and $
P_{FK}(R|s_{0})$ (knot). As expected, in the long wavelength limit
($ s_{0}\gg l$) the two distributions approach the Gaussian random
walk result, $P_{GRW}(R|s_{0})\propto R^{2} \exp[-3R^{2}/(4s_{0}l)]$
(see blue triangles). Since on very short length scales ($s_{0}<<l$)
all distribution functions approach the trivial limit
$\delta(R-s_{0}),$ the two distributions can only differ on
intermediate length scales ($s_{0}\approx l)$. \ This is indeed
confirmed by our simulation results, Fig. \ref{fig:RMS}. Notice that
in this regime the maximum of $P_{FK}(R|s_{0})$ is shifted to higher
values of $R$ than that of $P_{R}(R|s_{0})\ $indicating that typical
conformations of $1$-smooth curves are more compact than those of
$\infty$-smooth ones. The origin of the difference can be traced
back to the fact that the characteristic magnitude of the torsion of
a $1$-smooth curve is much larger than that of an $\infty $-smooth
one and, therefore, on length scales comparable to the persistence
length, the latter curves are confined to a plane while the former
have a three dimensional character.

\section{Discussion}

\label{sec:Discussion}We have demonstrated that the standard
continuum models of polymers including continuous Brownian random
walks, worm-like chains and ribbons, generate space curves that are
not sufficiently smooth to be described by the fundamental FS
equation of differential geometry. Examination of the corresponding
statistical ensembles shows that the dihedral angle $\Delta \varphi
$ between two successive binormals along the chain contour is
uniformly distributed in the interval $[-\pi ,\pi ]$ and we conclude
that both worm-like chains and centerlines of ribbons belong to the
class of freely rotating models, with divergent torsion and
discontinuous jumps of the normal to the curve. However, unlike
worm-like chains, ribbons have twist rigidity which means that the
rate of twist of the physical axes of the cross section remains
finite everywhere along the contour of the ribbon and guarantees
that the triad of unit vectors associated with the ribbon obeys the
generalized FS equation familiar from the differential geometry of
stripes. We compared some statistical properties of ensembles of
$1$-smooth and $\infty $-smooth curves generated by the ribbon and
the Fourier knot algorithms, respectively. \ We showed that in the
latter case the dihedral angle is peaked about $\Delta \varphi
\rightarrow 0$ and, therefore, typical configurations of Fourier
knots have finite torsion everywhere and can be described by the FS
equation. We also compared the distribution functions of the spatial
distance between two points along the contour of a ribbon and of a
Fourier knot. As expected, both distribution functions approach the
limiting Gaussian distribution for length scales much larger than
the persistence length, but are quite different on length scales
comparable to the persistence length.\newline Finally we would like
to stress that while the physical ensembles of conformations of
worm-like chains and ribbons are generated using the standard
methods of statistical physics (each conformation is weighted with
an appropriate Boltzmann factor, $\exp (-E/k_{b}T)$) , the ensemble
generated by the Fourier knot algorithm is a purely mathematical
construction and there is no elastic energy associated with
different conformations of Fourier knots. Nevertheless, the
observation of two persistence lengths reported in ref. \cite{Shay2}
and the present finding that Fourier knots have finite torsion,
suggest that the statistical properties of this mathematical
ensemble (notice that persistence lengths can be measured directly
from the ensemble of conformations of the space curves, just as is
done in AFM experiments \cite{experiment}) are quite similar to
those of a physical ensemble of conformations of polymers with both
bending and torsional rigidity. The detailed exploration of this
analogy is the subject of future work.
\acknowledgements This work was supported by a grant from the
US-Israel Binational Science Foundation.


\newpage


\end{document}